\documentstyle[12pt]{article}

\begin{document}

\baselineskip 0.7cm

\begin{titlepage}
\begin{flushright}
UT-945
\\
June, 2001
\end{flushright}

\vskip 1.35cm
\begin{center}
{\large \bf
Warped Compactification with a Four-Brane
}
\vskip 1.2cm
S.~Hayakawa and Izawa K.-I.
\vskip 0.4cm

{\it Department of Physics, University of Tokyo,\\
     Tokyo 113-0033, Japan}

\vskip 1.5cm

\abstract{
Warped compactification of the six-dimensional bulk with a negative
cosmological constant is realized with a 4-brane along with an abelian
gauge theory. No fine tuning of couplings are needed to obtain
the vanishing cosmological constant in four dimensions,
as is the case for the bulk with a positive cosmological constant.
}
\end{center}
\end{titlepage}

\setcounter{page}{2}

\section{Introduction}

Higher-dimensional approach to the cosmological constant
problem considered by Rubakov and Shaposhnikov
\cite{Rub}
is based on the idea that the vanishing cosmological constant
in four dimensions might be achieved through
warped compactification even in the presence of
a nonvanishing higher-dimensional cosmological constant.
This might be possible since effective four-dimensional flatness
is not necessarily in contradiction with highly curved spacetime
as a higher-dimensional space.

In a previous paper
\cite{Hayakawa:2000kk},
we investigated the case of the six-dimensional bulk with 
a positive cosmological constant and achieved the vanishing 
four-dimensional cosmological constant without metric singularity
by introducing an abelian gauge field
\cite{Wet}.
In this paper, we proceed to
the case of a negative bulk cosmological constant to get 
regular and compact extra dimensions. 
We are led to introduce a 4-brane in the bulk
\cite{Chacko:2000eb}
along with an abelian gauge field,
which is common to the case for a positive cosmological constant.

\section{The Model}

Let us assume an SO(2)-symmetric warped metric
as a six-dimensional background
\cite{Rub}-\cite{Chacko:2000eb}:
\begin{eqnarray}
 ds^2 = g_{MN}dx^M dx^N = \sigma(r){\bar g}_{\mu \nu}
    dx^{\mu}dx^{\nu} - dr^2 -\rho(r)d\theta^2,
\end{eqnarray}
where ${\bar g}_{\mu \nu}$ denote the four-dimensional
metric independent of $(r, \theta)$ with $0 \leq \theta < 2\pi$.
The action of the metric and an abelian gauge field
with a 4-brane at $r=r_1$ is given by
\begin{eqnarray}
 S = \int \! d^6 x \sqrt{-g}\left({1 \over 2} R  
     - {1 \over 4}F_{MN}F^{MN} - \Lambda \right)
     - \int \! d^6 x \sqrt{-g_5} \lambda \delta(r-r_1),
\end{eqnarray}
where $\Lambda < 0$ and $g_5$ denotes the determinant
of the induced metric on the 4-brane.

The gauge field equations of motion are obtained as
\begin{eqnarray}
 \partial_{M}(\sqrt{-g}F^{MN}) = 0.
\end{eqnarray}  
The background configuration with the SO(2) symmetry is
given by
\begin{eqnarray}
 A_{\mu} = A_{r} = 0, \, \, A_{\theta} = a(r),
\end{eqnarray}
which yield the following field strength:
\begin{eqnarray}
 F^{r \theta} = {B \over {\sigma^2 \sqrt{\rho}}},
\end{eqnarray}
where $B$ is an integration constant.
With this gauge field configuration, the Einstein equations
lead to
\begin{eqnarray}
 & & {3 \over 2}{\sigma'' \over \sigma}
     +{3 \over 4}{\sigma' \over \sigma}{\rho' \over \rho}
     -{1 \over 4}{\rho'^2 \over \rho^2}
     +{1 \over 2}{\rho'' \over \rho} - {\Lambda_4 \over \sigma}
     = -{B^2 \over 2\sigma^4} - \Lambda - \lambda \delta(r-r_1),
 \\
 & & {3 \over 2}{\sigma'^2 \over \sigma^2}
     +{\sigma' \over \sigma}{\rho' \over \rho}
     -{2\Lambda_4 \over \sigma}
     = {B^2 \over 2\sigma^4}-\Lambda,
 \\
 & & 2{\sigma'' \over \sigma}+{1 \over 2}{\sigma'^2 \over \sigma^2}
     -{2\Lambda_4 \over \sigma}
     = {B^2 \over 2\sigma^4}-\Lambda - \lambda \delta(r-r_1),
\end{eqnarray}
where the prime denotes differentiation with respect to $r$
and we have used four-dimensional Einstein equations 
for the metric ${\bar g}_{\mu \nu}$
with the cosmological constant $\Lambda_4$,
which comes out as an integration constant.

We get the junction conditions
across the 4-brane from the equations of motion:
\begin{eqnarray}
 & & \sigma'(r_1 +0) - \sigma'(r_1 -0)  = -{1 \over 2}\lambda \sigma(r_1),
     \label{juns} \\
 & & \rho'(r_1 +0) - \rho'(r_1 -0) = -{1 \over 2}\lambda \rho(r_1). 
     \label{junr}
\end{eqnarray}
In the regions $r < r_1$ and $r > r_1$, the bulk Einstein equations are
reduced to an equation of motion
\begin{eqnarray}
 & & z'' = - {{\partial V(z)} \over {\partial z}}; \quad
     V(z) = {25 \over 96}B^2 z^{-6/5} + {5 \over 16}\Lambda z^2 
          - {25 \over 24}\Lambda_4 z^{6/5},
     \label{eq}\\
 & & \sigma = z^{4/5}, \quad \rho = C^{-2}_{\mp}z'^2 z^{-6/5},
     \label{eqa} 
\end{eqnarray}
where $C_{\mp}$ denote an integration constant for each region.
Note that this equation describes the motion of a particle 
with the position $z$ at the
time $r$ in a potential $V(z)$.

\section{The Solutions}

Now we seek regular metric solutions
in the desirable case of $\Lambda_4 =0$.
The junction conditions
Eqs.(\ref{juns}) and (\ref{junr})
can be expressed in terms of $z$
under $\lambda \neq 0$:
\begin{eqnarray}
 & & z'_{+} - z'_{-} = -{5 \over 8}\lambda z_1,
      \label{junz1}\\
 & & z'_{+}z'_{-} 
      = - {5 \over 16}B^2 z_1^{-{6/5}} + {5 \over 8}\Lambda z_1^2
      \label{junz2},
\end{eqnarray}
where we have denoted $z'_\pm=z'(r_1 \pm 0)$
and $z_1=z(r_1)$
(which may be normalized to one without loss of generality).
Eqs.(\ref{junz1}) and (\ref{junz2}) can be solved for $z'_{\pm}$,
provided $5\lambda^2 + 32\Lambda \geq 0$,
with $z'_{-} > 0$ and $z'_{+} < 0$.
This change of the velocity $z'$ from positive to negative
is crucial for compactification of the extra dimensions,
which is achieved by means of the 4-brane with $\lambda > 0$.

The equation of motion Eq.(\ref{eq})
implies $z' = 0$ at $r = r_0, \bar{r}$
with $z' \neq 0$ for $r_0 < r <r_1$ and $r_1 < r < \bar{r}$.
The `energy' conservations as the motion of the position $z$
in the `time' regions $r_0 \leq r < r_1$ and $r_1 < r \leq \bar{r}$ yield
\begin{eqnarray}
 & & {25 \over 96}B^2 z^{-6/5}_0 + {5 \over 16}\Lambda z^2_0 
     = {1 \over 2}z'^2_{-} 
     + {25 \over 96}B^2 z^{-6/5}_1 + {5 \over 16}\Lambda z^2_1,
      \label{z0}\\
 & & {25 \over 96}B^2 \bar{z}^{-6/5} + {5 \over 16}\Lambda \bar{z}^2 
     = {1 \over 2}z'^2_{+} 
     + {25 \over 96}B^2 z^{-6/5}_1 + {5 \over 16}\Lambda z^2_1,
      \label{barz}
\end{eqnarray}
respectively.
Here we have denoted $z_0=z(r_0)$, ${\bar z}=z(\bar{r})$,
whose values are determined by these equations.

Continuity of $\rho$ across the 4-brane imposes a condition
\begin{eqnarray}
   {C_{-}^{-2}}{z'^2_{-}} = {C_{+}^{-2}}{z'^2_{+}}
      \label{con}
\end{eqnarray}
from Eq.(\ref{eqa}).
For regularity of the higher-dimensional metric, we have conditions
\begin{equation}
 (\sqrt{\rho})'(r_0) = -(\sqrt{\rho})'(\bar{r}) =1,
\end{equation}
that is,
\begin{eqnarray}
 & & C_{-}z^{3/5}_0 = {5 \over 16}B^2 z^{-11/5}_0 - {5 \over 8}\Lambda z_0,
      \label{bz0} \\
 & & C_{+}\bar{z}^{3/5} = {5 \over 16}B^2 \bar{z}^{-11/5}
                        - {5 \over 8} \Lambda \bar{z}.
      \label{bbz}
\end{eqnarray}
The above three conditions can be solved for
three integration constants $B$ and $C_{\mp}$,
which yield a reflection symmetric metric with respect to
the 4-brane.

Namely, the effective four-dimensional cosmological constant
$\Lambda_4$ can vanish for generic values of
$\Lambda < 0$ and $\lambda > 0$ with the extra dimensions compactified.

We note that, in addition to the 4-brane at $r=r_1$,
3-branes may be introduced at $r=r_0$ and/or $r=\bar{r}$
straightforwardly.
For example, if we put a 3-brane with the tension 
$\lambda_0$ at $r=r_0$, the regularity
condition Eq.(\ref{bz0}) is replaced by 
\cite{Hayakawa:2000kk} 
\begin{eqnarray}
 C_{-}\bigl( 1 - {\lambda_0 \over 2\pi}\bigr) z^{3/5}_0 =
     {5 \over 16}B^2 z^{-11/5}_0 - {5 \over 8} \Lambda z_0,
\end{eqnarray}
which also allows the full equations of motion
to be satisfied without singularity.

\section{Conclusion}

We have obtained the regular metric with compact extra dimensions
in the case of the negative bulk cosmological constant
in six dimensions.
The four-dimensional
cosmological constant $\Lambda_4$ can vanish without fine tuning of
Lagrangian parameters $\Lambda < 0, \lambda > 0$
(and the tensions of 3-branes, if present).

We note that regular and compact metrics are also present under
$\Lambda_4 \neq 0$, as is the case for a positive bulk
cosmological constant
\cite{Hayakawa:2000kk}.
Namely, $\Lambda_4$ is none other than an integration constant
to be determined by boundary conditions,
which might be supplied by some other sector
to be added in the model
\cite{Iza}.

\end{document}